 \documentclass[final,5p,times,twocolumn,compress]{elsarticle}

\usepackage{amssymb}
\usepackage{lipsum}
\usepackage{booktabs}
\usepackage[colorlinks=true,citecolor=blue,linkcolor=magenta]{hyperref}
\journal{Physics Letters B}

\begin{document}

\begin{frontmatter}



\title{Qubit-efficient variational algorithm for nuclear structure}


\author[first]{Chandan Sarma}
\author[first]{P. D. Stevenson}
\affiliation[first]{organization={School of Mathematics and Physics, University of Surrey},%Department and Organization
            addressline={}, 
            city={Guildford},
            postcode={GU2 7XH}, 
            state={Surrey},
            country={United Kingdom}}

\begin{abstract}
In this work, we compare three qubit-mapping strategies to study the structure of the nuclear ground state within the  shell model description employing the Variational Quantum Eigensolver (VQE) approach. Although the initial point for different mappings is a Hamiltonian matrix in many-body particle basis or Slater determinant (SD) basis, the structure of the trial wavefunction and resource counts are different for each mapping. These three mappings are tested for a mid $p$-shell nucleus $^{10}$B and compared the quantum resources required to find the ground state for each mapping. Further, we extend the qubit-efficient mapping to study the ground state of one more mid $p$-shell nucleus $^{12}$C. We run circuits up to 26-qubits representing their ground states on a noisy simulator (IBM’s FakeFez backend) and quantum hardware ($ibm\_fez$). 
{\color{black}
Following zero-noise extrapolation (ZNE), the mean ground-state energies obtained from hardware are systematically improved, in the sense that they approach the exact reference results.  Uncertainties associated with the ZNE vary across the mappings, which we attribute to varying circuit complexity.}
%{\color{black}Following zero-noise extrapolation (ZNE), the mean ground-state energy obtained from hardware for \(^{10}\mathrm{B}\) using the SD mapping differs from the reference energy by 0.21\%. The corresponding deviations for the pnSD and cSD mappings are 8.88\% and 3.37\%, respectively. For \(^{12}\mathrm{C}\), the mean ZNE-extrapolated ground-state energy obtained using the cSD mapping differs from the reference energy by 6.82\%. These percentages quantify the residual deviations of the mean ZNE estimates from the reference energies, and the total uncertainty of the extrapolation for SD and pnSD mapping is large compared to the cSD mapping.}
%The best post-error mitigated results from the hardware for $^{10}$B ground state is obtained following SD to qubit mapping with a percent error of 0.21 \%. The percent errors for the same state following cSD and pnSD mapping are 3.37 and 8.88 \%, respectively. On the other hand, following the cSD mapping, the post-error mitigated ground state energy of $^{12}$C is 6.82 \% away from the exact result. 
We further evaluate the fidelity of the VQE wavefunctions obtained from hardware with respect to the shell model wavefunctions for the cSD mapping. This cSD mapping can be useful for scaling the VQE algorithm for complex nuclei across different mass regions in terms of qubit efficiency. 

\end{abstract}




\end{frontmatter}

%\tableofcontents



\section{Introduction}
\label{introduction}
Quantum many-body systems arising in atomic nuclei, which involve anywhere from a handful to several hundred particles, offer an ideal setting for exploring current and future applications of quantum computing platforms. There are many computationally classical methods available to understand the quantum structure of atomic nuclei, starting from the conventional nuclear shell model \cite{SM1, SM2, shellmodelmdpi, CORAGGIO2024104079, Liu_2023} to recent developments in \textit{ab initio} methods \cite{NCSM, ab_initio, Hu2022Ab, Sarma_2023, Sarma2025, QMC1, QMC2, SCGF1, SCGF2, CC}. However, all these methods ultimately face the combinatorially increasing size of the Hilbert space with the number of nucleons, which means that many nuclear systems remain beyond the reach of these methods on classical computers.

Quantum computing has the potential to make large nuclear shell model calculations feasible by leveraging the exponential growth of multi‑qubit Hilbert spaces and the ability to efficiently represent strongly entangled states. As a result, the shell model has become a natural platform for quantum algorithms to study the structure of atomic nuclei. Its $M$-scheme formulation \cite{deshalit_talmi,whitehead_numerical_1972} maps single particle states directly onto qubits and provides a unified framework that can address both simple benchmark problems and realistic many‑body systems \cite{QC_rev,kiss_quantum_2022,stetcu,sarma_prediction_2023,perez-obiol_nuclear_2023,bhoy_shell-model_2024, 17n2-xh6k,nifeeya,costa_quantum_2025, sarma_low-circuit-depth_2026, kimura_prc, kimura_prr, COSTA2026140042}.
{\color{black}There are also some efforts to perform \textit{ab inito} calculations using quantum computation methods \cite{Mumma_2024, gqdy-dvps, jiang2026textitabinitioquantumsimulations, 1spg-5ld3}.}
Many of these works focus on a hybrid algorithm, called the variational quantum eigensolver (VQE), based on parametrized wavefunctions that aim to reproduce the lowest-energy state corresponding to a Hamiltonian by optimizing the variational parameters. Some adjacent works focus on excited states as well while following the VQE strategy \cite{ashutosh, Denis_plb, maheshwari_single-step_2026,hobday_variance_2025}. Designing an appropriate parametrized wavefunction for a given system is crucial to the success of the VQE algorithm. Among the different ways to design a trial wavefunction for VQE, the unitary coupled cluster (UCC) \cite{10.1039/d1cs00932j} and the adaptive derivative-assembled pseudo-Trotter VQE (ADAPT-VQE) \cite{Grimsley2019} are two of the most widely explored. Both types of trial wavefunctions are usually based on a single-particle state to qubit mapping. {\color{black}Apart from these hybrid algorithms, some methods employ projection techniques \cite{PhysRevC.108.L031306, li_deep_2024}, while others rely on quantum annealing protocols \cite{SciPostPhys.19.2.062}. There are also some efforts to look beyond the 
{\color{black}noisy intermediate-scale quantum (NISQ) }
era \cite{Preskill2018quantumcomputingin} towards the early Fault-Tolerant era in the context of nuclear structure \cite{gibbs2026lowtcountpreparationnuclear, PsiQ, yoshida2026nuclearmanybodysystemsbenchmarks}.}

In our previous work, we considered a many-body configuration-to-qubit mapping within the VQE formalism that led to low-depth circuits for a selected set of low- and high-mass nuclei near closed shell configurations \cite{sarma_low-circuit-depth_2026} compared to the UCC type trial wavefunctions based on single particle state to qubit mapping. The trade-off for our low-depth circuits was an increased number of qubits, which limits their application to a small number of nuclei in different mass regions before implementing any truncation scheme on the allowed many-body states. In this work, we address this issue of requiring a larger number of qubits by modifying the earlier mapping technique. Here, we introduce two different mapping techniques, both of which are based on mapping problem-specific many-particle states or SDs to the qubit Hilbert space. We compare the resource counts and performance of these three different mappings, and also compare the advantages of each mapping over the others. 
%{\color{black}In Refs. \cite{kimura_prr, COSTA2026140042}, hard-core boson or quasiparticle encoding was introduced and used for nuclear systems, which helps in avoiding the nonlocal Pauli strings and simplifications in quantum circuits with a similar motivation to the present work.}
We use the mid-$p$-shell nucleus 
$^{10}$B as a test case to compare three mapping techniques for calculating its ground state within the nuclear shell model. The choice of $^{10}$B is ideal as it represents the highest $M$-scheme dimension of 84 among the $p$-shell nuclei, which require deep circuits following UCC and ADAPT-VQE type ansatzes \cite{17n2-xh6k}. We further demonstrate a qubit-efficient mapping to study another mid-$p$-shell nucleus, $^{12}$C, which is computationally expensive {\color{black} for the other two mapping techniques. In Refs. \cite{kimura_prr, COSTA2026140042}, hard-core boson and quasiparticle encodings were introduced and applied to nuclear systems, with the aim of avoiding nonlocal Pauli strings and simplifying quantum circuits, in a manner closely related to the motivation of the present work.} 

%We consider a mid-$p$-shell nucleus $^{10}$B as our test case, which is an ideal case to test three different mapping techniques to evaluate the ground state of this nucleus within the nuclear shell model formalism.

\section{Formalism}
\begin{figure}
    \centering
    \includegraphics[width=0.47\textwidth, trim=1cm 1cm 1cm 3cm, clip]{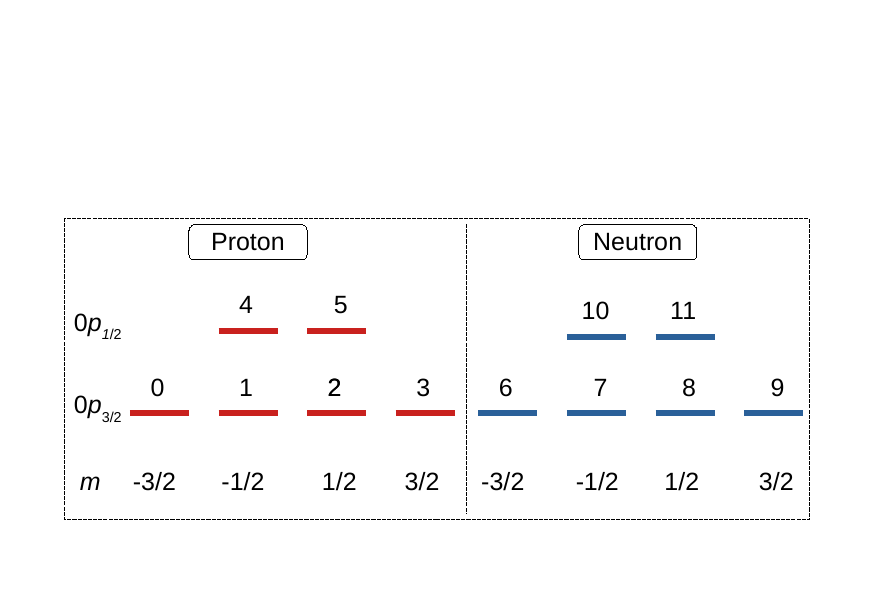}
    \caption{The 12 single particle states representing the $p$-shell is shown. The quantum numbers, $n, l, j, m$, are given in usual nuclear shell model notation.}
    \label{pshell}
\end{figure}

%%\label{}
% \lipsum[1]
Within the shell model formalism, $^{10}$B can be defined as a mid-$p$ shell nucleus having six valence nucleons above the $^4$He core. We use the well-known effective shell model interaction CKpot \cite{ckpot} for this work, which defines effective interactions among the nucleons occupying the 0$p_{3/2}$ and 0$p_{1/2}$ orbitals. Within the $M$-scheme formulation, many-particle states for the $p$-shell nuclei can be constructed from 12 single-particle states shown in Fig. \ref{pshell}. The spin-parity of the ground state of $^{10}$B, the nucleus of our interest, is 3$^+$. This state can be constructed by a linear combination of  26 many-particle states with $M$ = 3.  It is also possible to construct many-particle states for $M$ = 0, $\pm$1, and $\pm$2 representing the same state, which require more than 26 many-particle states. We choose $M$ = 3 many-particle basis for $^{10}$B ground state  The classical shell model approach involves constructing a Hamiltonian matrix in the many-particle basis and diagonalising it for eigenvalues. We follow the same strategy as the shell model formalism as far as the Hamiltonian construction in many-particle or Slater determinant (SD) basis, and then we turn to construct suitable Hamiltonians and design quantum circuits for the VQE algorithm, following three different strategies for mapping many-particle states into qubit Hilbert space. 

\begin{table}[t]
\centering
\small
\caption{Qubit assignments for the ground state of $^{10}$B under SD and pnSD mappings. In the SD mapping, 26 many-particle states correspond to 26 qubits. In the pnSD mapping, the first ten qubits represent proton SDs, and the next ten represent neutron SDs.}
    \begin{tabular}{ccc}
    \hline
        Qubits &  SD mapping& pnSD mapping \\ %& Compact SD mapping \\
    \hline    
         0 & |0, 2, 3; 8, 9, 11$\rangle$ & |0, 2, 3$\rangle$ \\%& |00000$\rangle$ \\
         1 & |0, 3, 5; 8, 9, 11$\rangle$ & |0, 3, 5$\rangle$ \\%& |00001$\rangle$ \\
         2 & |1, 2, 3; 7, 8, 9$\rangle$ & |1, 2, 3$\rangle$ \\%& |00010$\rangle$ \\
         3 & |1, 2, 3; 7, 9, 11$\rangle$ & |1, 2, 5$\rangle$ \\%& |00011$\rangle$ \\
         4 & |1, 2, 3; 8, 9, 10$\rangle$ & |1, 3, 4$\rangle$ \\%& |00100$\rangle$ \\
         5 & |1, 2, 3; 9, 10, 11$\rangle$ & |1, 3, 5$\rangle$ \\%& |00101$\rangle$ \\
         6 & |1, 2, 5; 8, 9, 11$\rangle$ & |2, 3, 4$\rangle$ \\%& |00110$\rangle$ \\
         7 & |1, 3, 4; 8, 9, 11$\rangle$ & |2, 3, 5$\rangle$ \\%& |00111$\rangle$ \\
         8 & |1, 3, 5; 7, 8, 9$\rangle$ & |2, 4, 5$\rangle$ \\%& |01000$\rangle$ \\
         9 & |1, 3, 5; 7, 9, 11$\rangle$ & |3, 4, 5$\rangle$ \\%& |01001$\rangle$ \\
         10 & |1, 3, 5; 8, 9, 10$\rangle$ & |8, 9, 11$\rangle$ \\%& |01010$\rangle$ \\
         11 & |1, 3, 5; 9, 10, 11$\rangle$ & |7, 8, 9$\rangle$ \\%& |01011$\rangle$ \\
         12 & |2, 3, 4; 7, 8, 9$\rangle$ & |7, 9, 11$\rangle$ \\%& |01100$\rangle$ \\
         13 & |2, 3, 4; 7, 9, 11$\rangle$ & |8, 9, 10$\rangle$ \\%& |01101$\rangle$ \\
         14 & |2, 3, 4; 8, 9, 10$\rangle$ & |9, 10, 11$\rangle$\\% & |01110$\rangle$ \\
         15 & |2, 3, 4; 9, 10, 11$\rangle$ & |6, 8, 9$\rangle$ \\%& |01111$\rangle$ \\
         16 & |2, 3, 5; 6, 8, 9$\rangle$ & |6, 9, 11$\rangle$\\% & |10000$\rangle$ \\
         17 & |2, 3, 5; 6, 9, 11$\rangle$ & |7, 8, 11$\rangle$ \\%& |10001$\rangle$ \\
         18 & |2, 3, 5; 7, 8, 11$\rangle$ & |7, 9, 10$\rangle$ \\% & |10010$\rangle$ \\
         19 & |2, 3, 5; 7, 9, 10$\rangle$ & |8, 10, 11$\rangle$ \\%& |10011$\rangle$ \\
         20 & |2, 3, 5; 8, 10, 11$\rangle$ & - \\%& |10100$\rangle$ \\
         21 & |2, 4, 5; 8, 9, 11$\rangle$ & - \\%& |10101$\rangle$ \\
         22 & |3, 4, 5; 7, 8, 9$\rangle$ & - \\%& |10110$\rangle$ \\
         23 & |3, 4, 5; 7, 9, 11$\rangle$ & - \\%& |10111$\rangle$ \\
         24 & |3, 4, 5; 8, 9, 10$\rangle$ & - \\%& |11000$\rangle$ \\
         25 & |3, 4, 5; 9, 10, 11$\rangle$ & - \\%& |11001$\rangle$ \\
    \end{tabular}
    \label{qubit_assign}
\end{table}

\subsection{Slater determinant (SD) mapping}
The Hamiltonian matrix constructed for the ground state of $^{10}$B within the SDs with $M$ = 3 is 26-dimensional, where each row and column represents a particular Slater determinant (SD). Within this mapping, each possible SD is considered as a qubit, as shown in Table \ref{qubit_assign}, and the 26-dimensional Hamiltonian matrix can be converted into Pauli strings as follows:

\begin{equation}
\label{eq1}
H_{qubit}^{SD} = \sum_{m} H_{mm} \frac{(I_m -Z_m)}{2} + \sum_{m < n} H_{mn} \frac{(X_m X_n + Y_m Y_n)}{2}.
\end{equation}

Here, $H_{mm}$ and $H_{mn}$ are the diagonal and off-diagonal many-particle matrix elements.
It should be noted that this simple form of the Hamiltonian in Pauli string form is possible because each qubit maps to a full SD, which already accounts for antisymmetry.

The 26-qubit trial wavefunction for this mapping, having 25 adjustable parameters, is shown in the first panel of Fig. \ref{ansatze}. This type of mapping was earlier explored in \cite{sarma_low-circuit-depth_2026} up to a 29-qubit problem. The initial state $|1, 0,..., 0\rangle$ obtained by using an $X$ gate to qubit 0 represents the 6-particle SD $|0, 2, 3; 8, 9, 11\rangle$ as shown in Table \ref{qubit_assign}. The staircase of $G^1$ operators represents single-excitation gates based on Givens rotation that can be further broken down into a set of 
{\color{black}Hadamard ($H$), controlled-NOT ($CNOT$), and single-qubit rotation gate about the Y-axis ($R_Y$) }
as shown in Fig. \ref{Double_ex}.

\subsection{Proton neutron Slater determinant (pnSD) mapping}
It is seen that the SD mapping requires the same number of qubits as the number of SDs for a particular nuclear system. One way to reduce the required number of qubits is by splitting the allowed SDs into proton SD (pSD) and neutron SD (nSD) components and assigning unique pSDs and nSDs as qubits. For the case of $^{10}$B (3$^+$), the 26 SDs can be constructed by using 10 unique pSDs and the same number of nSDs, which are shown in Table \ref{qubit_assign}. Hence, within this qubit mapping, $^{10}$B (3$^+$) is defined as a 20-qubit problem, and the Hamiltonian can be represented as

\begin{equation}
\label{eq2}
H^{pnSD} =  \sum_{(p, n), (p', n')} V_{pnp'n'} \hat{a}^\dagger_p \hat{a}^\dagger_n \hat{a}_{p'} \hat{a}_{n'}.
\end{equation}

Here, ($p,n$) represents pSDs and nSDs and $V_{p,n,p',n'}$ represents the many-particle matrix elements of the Hamiltonian matrix. The operator $\hat{a}^\dagger_p \hat{a}^\dagger_n$ represents the creation of a possible many-particle state. 
The Hamiltonian in Eq. \ref{eq2} can be converted into Pauli strings by following the standard Jordan-Wigner mapping \cite{JW}, in which antisymmetry is enforced, as follows

\begin{eqnarray}
\hat{a}^\dagger_k = \frac{1}{2}  \left ( \prod_{j = 0}^{k-1} -Z_j \right)(X_k -iY_k), \label{eq3}\\
\hat{a}_k = \frac{1}{2}  \left ( \prod_{j = 0}^{k-1} -Z_j \right)(X_k + iY_k). \label{eq4} 
\end{eqnarray}

The trial wavefunction for this mapping is shown in the second panel of Fig. \ref{ansatze}, having 25 adjustable parameters, the same as the SD mapping case. The initial state |0, 10$\rangle$, which is obtained by applying $X$ gates to qubits 0 and 10, represents the 6-nucleon SD |0, 2, 3; 8, 9, 11$\rangle$, where |0, 2, 3$\rangle$ is the pSD represented by qubit 0 and |8, 9, 11$\rangle$ is the corresponding nSD represented by qubit 10. The other 25 SDs are obtained by applying layers of controlled single-excitation gates ($CG^1$) and double excitation gates ($G^2$). A total of 23 $CG^1$ gates and two $G^2$ gates are used in constructing the trial wavefunction; only a part is shown in Fig. \ref{ansatze}. The $CG^1$ gate can be decomposed into a sequence of $R_Y$ and $CNOT$ gates, while the $G^2$ gates can be decomposed into sequences of $H$, $CNOT$, and $R_Y$ gates, as shown in \ref{Double_ex}. 

\subsection{Compact Slater determinant (cSD) mapping}
While the SD mapping yields a simple, low-depth trial wavefunction at the expense of more qubits, the pnSD mapping requires fewer qubits at the expense of a deeper circuit. In a third mapping, we address the disadvantages of the earlier two mappings. First, we convert the 26-dimensional Hamiltonian matrix into a 32-dimensional matrix (the nearest dimension of the qubit Hilbert space, represented by $2^n$) by adding six rows and columns of all zeros. Now, we perform a Pauli conversion of the new 32-dimensional matrix using Qiskit to obtain a 5-qubit Hamiltonian in Pauli string form. 

\begin{figure*}
    \centering
    \includegraphics[width=0.55\linewidth]{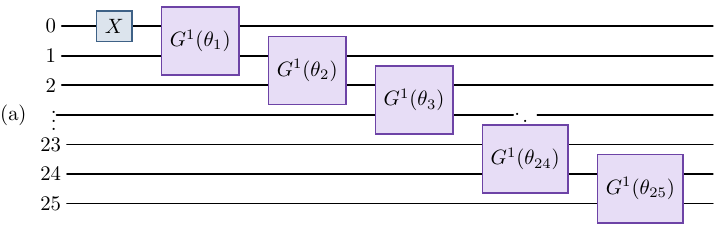}\\
    \includegraphics[width=0.85\linewidth]{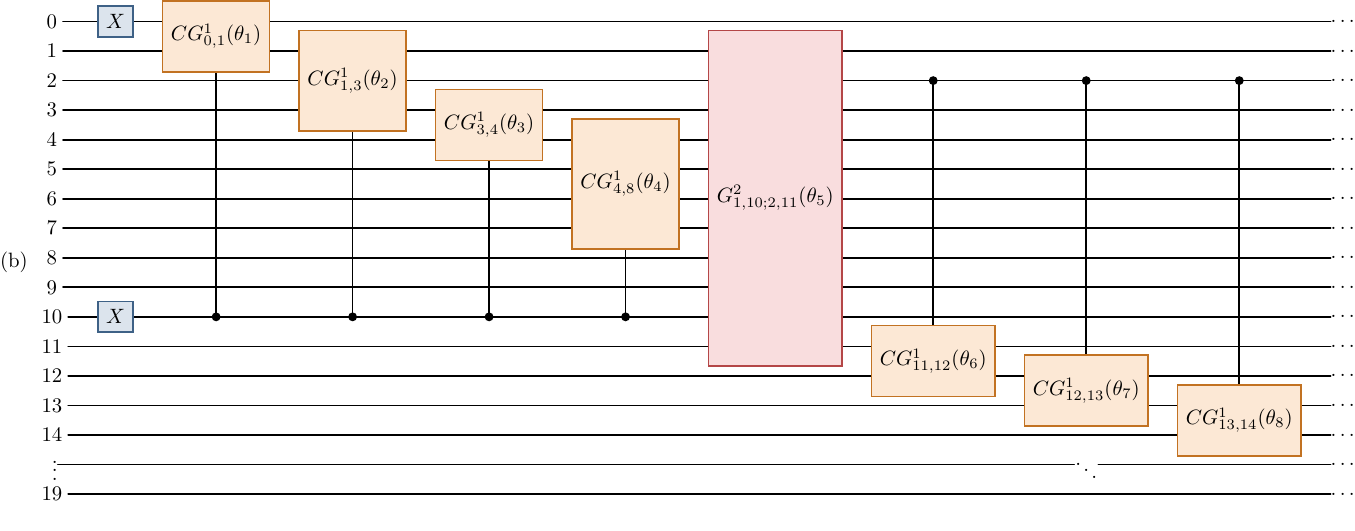}\\
    \includegraphics[width=0.75\linewidth]{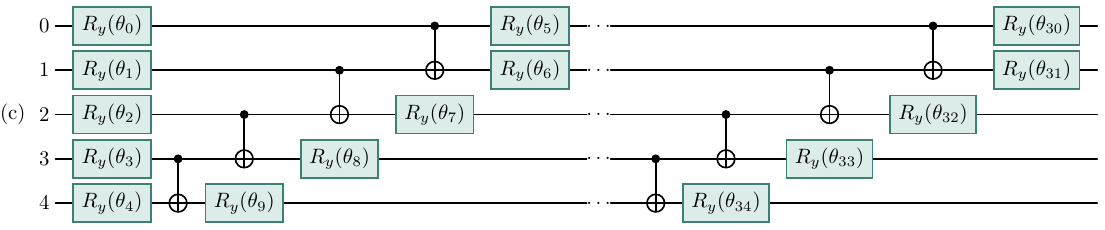}
    \caption{(a) 26-qubit circuit for $^{10}$B ground state following SD mapping, (b) 20-qubit circuit for the same nucleus considering pnSD mapping, (c) 5-qubit circuit for the ground state of $^{10}$B using cSD .}
    \label{ansatze}
\end{figure*}

The trial wavefunction for cSD mapping is shown in the last panel of Fig. \ref{ansatze}, consisting of alternating layers of $R_Y$ and $CNOT$ gates. One disadvantage of this hardware-efficient ansatz is the requirement of a larger number of parameters compared to the earlier two mappings. For the case of $^{10}$B ground state, six layers of  $R_Y$ and $CNOT$ gates were required to exactly converge to the ground state binding energy having 35 variational parameters. This particular mapping is further extended to define the ground state (0$^+$) of another mid-$p$-shell nucleus, $^{12}$C, where the $M$-scheme dimension of the Hamiltonian is 51. The SD and pnSD mapping require 51 and 30 qubits to define the ground state of this nucleus, respectively, placing them at or beyond the capabilities of a typical classical computer when using a statevector simulator. Hence we choose to follow cSD mapping only for this case where, we convert the 51-dimensional Hamiltonian to a 64-dimensional matrix like the $^{10}$B case. We then perform a conversion of the 64-dimensional matrix to obtain a 6-qubit Hamiltonian in Pauli string form. The structure of the trial wavefunction for $^{12}$C is the same as that of $^{10}$B, as shown in the last panel of Fig. \ref{ansatze} with one additional qubit. For this case, 10 layers of $R_Y$ and $CNOT$ gates were required to exactly converge to the ground-state binding energy, having 66 variational parameters.
{\color{black}For the two nuclei considered in this work, the required number of parameters to converge to the exact ground-state binding energy is roughly 1.3 times the number of SDs. These numbers can vary based on the many-body structure of the system under consideration, and it would be interesting to look at its possible link to the underlying entanglement structure \cite{magic_prc}. For the $sd$-shell nuclei, the UCC or ADAPT-VQE requires 24 qubits based on single-particle state-to-qubit mapping. In contrast, cSD mapping represents a qubit-efficient variant requiring up to 17 qubits for the mid-$sd$-shell nucleus, $^{28}$Si, having an $M$-scheme dimension of 93710.}

\begin{figure}[t]
    % \centering
    \includegraphics[width=0.9\linewidth]{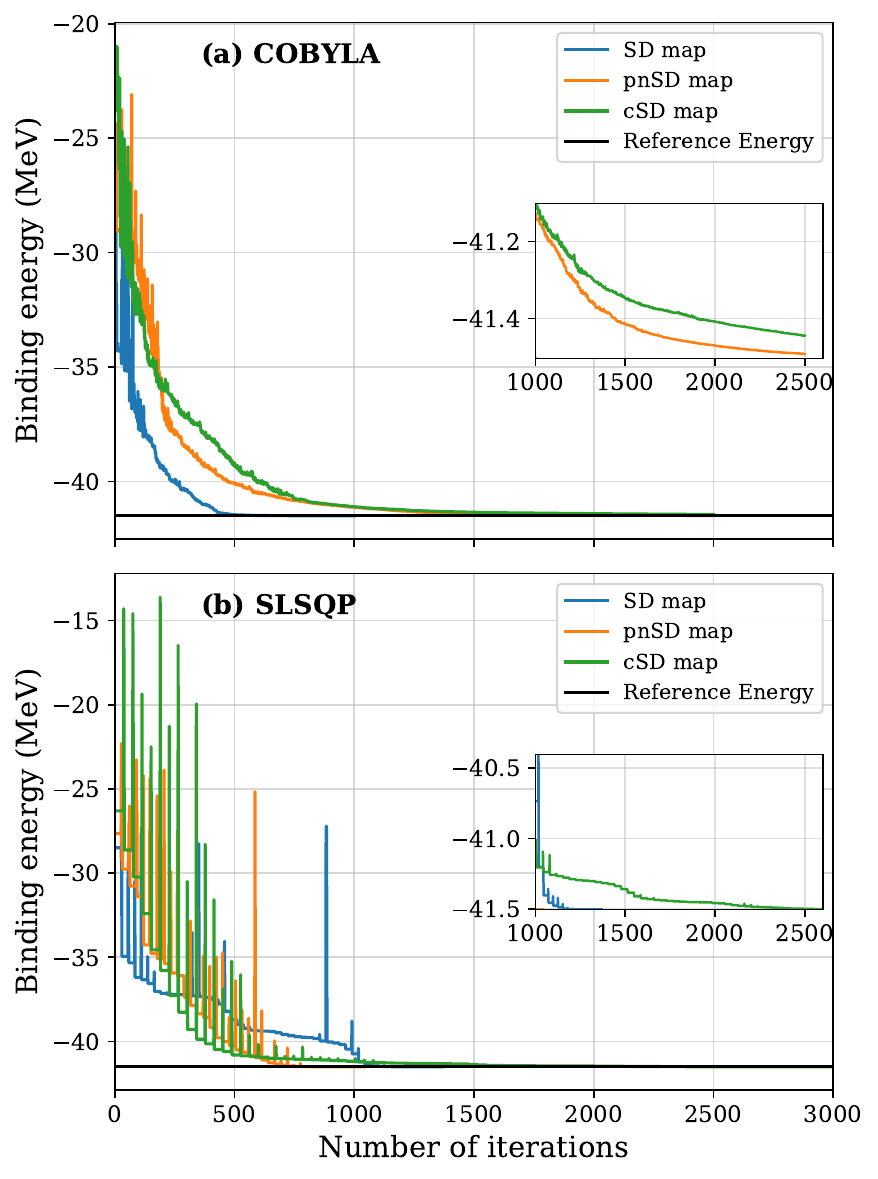}
\caption{Convergence of $^{10}$B ground state binding energies with the number of iterations using (a) COBYLA optimizer and (b) SLSQP optimizer.}
    \label{10B_vqe}
\end{figure}  

\begin{table*}
\caption{The ground state spins and parities of $^{10}$B and $^{12}$C, and their ground state reference (“Ref.”) energies (in MeV) from shell model calculations are shown along with the required resource counts to simulate $^{10}$B ground state using three different mappings. In addition, resource counts for $^{12}$C ground state are also shown following qubit-efficient cSD mapping.}
\centering
\scriptsize
\setlength{\tabcolsep}{12pt}
\begin{tabular}{cccccccccc}
\toprule
Nucleus ($J^\pi$) & Mapping & Qubits & Params & Pauli & Comm. Pauli& 1Q & 2Q & Depth & Ref. Energy\\
\midrule
$^{10}$B (3$^+$) & SD (Noiseless sim.) & 26 & 25 & 451 & 4 & 101 & 50 & 102 & -41.504 \\
 & SD (Noisy sim.) & 26 & 25 & 451 & 4 & 391 & 50 & 229 &  \\
  & SD (Hardware) & 26 & 25 & 451 & 4 & 364 & 50 & 211 &  \\
\midrule
 & pnSD (Noiseless sim.) & 20 & 25 & 1271 & 530 & 122 & 166 & 258 &  \\
 & pnSD (Noisy sim.) & 20 & 25 & 1271 & 530 & 1071 & 305 & 746 &  \\
  & pnSD (Hardware) & 20 & 25 & 1271 & 530 & 1143 & 314 & 762 &  \\
\midrule
 & cSD (Noiseless sim.) & 5 & 35 & 528 & 122 & 35 & 24 & 21 &  \\
 & cSD (Noisy sim.) & 5 & 35 & 528 & 122 & 169 & 24 & 61  \\
  & cSD (Hardware) & 5 & 35 & 528 & 122 & 169 & 24 & 61 &  \\
 \midrule
 \midrule
 $^{12}$C (0$^+$) & cSD (Noiseless sim.) & 6 & 66 & 2072 & 365 & 66 & 50 & 34 & -71.045   \\
 & cSD (Noisy sim.) & 6 & 66 & 2072 & 365 & 336 & 50 & 97 &    \\
 & cSD (Hardware) & 6 & 66 & 2072 & 365 &  336 & 50 & 97 &    \\
\bottomrule
\end{tabular}
\label{rescource}
\end{table*}

\begin{table*}
\caption{Numerical results from noisy simulation and hardware are shown.}
\centering
\scriptsize
\setlength{\tabcolsep}{12pt}
\begin{tabular}{cccccc}
\hline
Nucleus ($J^\pi$) & Mapping & Noise 1 & Noise 3 & ZNE \\
\hline
$^{10}$B (3$^+$) & SD (Noisy sim.) & -50.422$\pm$2.506 & -53.258$\pm$3.194 & -49.004$\pm$4.084\\
  & SD (Hardware) & -73.098$\pm$2.827 & -136.108$\pm$3.924 &  -41.593$\pm$4.672 \\
\hline
$^{10}$B (3$^+$) & pnSD (Noisy sim.) & -42.130$\pm$2.877 & -55.303$\pm$4.573 &  -35.544$\pm$4.884 \\
  & pnSD (Hardware) & -54.771$\pm$8.196 & -73.937$\pm$14.728 & -45.188$\pm$14.331 \\
\hline
$^{10}$B (3$^+$) & cSD (Noisy sim.) & -38.913$\pm$0.330 & -36.798$\pm$0.162  & -39.970$\pm$0.502 \\
  & cSD (Hardware) &  -39.199$\pm$0.289 & -37.386$\pm$0.218 & -40.106$\pm$0.447   \\
\hline
\hline
$^{12}$C (0$^+$) & cSD (Noisy sim.) & -64.225$\pm$0.339 & -57.618$\pm$0.555 & -67.528$\pm$0.579 \\
  & cSD (Hardware) & -63.547$\pm$2.256 &-58.236$\pm$1.590 &  -66.202$\pm$3.476  \\
\hline
\end{tabular}
\label{results}
\end{table*}

\section{Results}
In this work, the parameters of the trial wavefunctions are determined using a standard VQE procedure, where the initial VQE optimization is carried out in a noiseless environment using the Statevector simulator from Qiskit. Once the parameters are obtained from the noiseless simulation, the resulting circuits with these fixed angles are executed on a noisy simulator and real quantum hardware. Table \ref{rescource} shows the ground-state spin-parities of $^{10}$B and $^{12}$C, together with their reference ground-state energies (in MeV) obtained from shell-model calculations. The resource requirements for simulating the $^{10}$B ground state using three different mappings are shown
{\color{black}in terms of number of qubits, number of parameters, number of Pauli terms in the qubit Hamiltonian, number of commuting Pauli groups, one- and two-qubit gates and depth of the circuits.
The resource requirements for the $^{12}$C ground state following the cSD mapping are also shown in the table.}
For each mapping, required resources are provided in three rows corresponding to the noiseless simulation (Noiseless sim.), the noisy simulation using IBM's FakeFez backend (Noisy sim.), and running them on $ibm\_fez$ quantum computer (Hardware).

\subsection{Noiseless simulation}

To determine the optimized ground state, we consider the original ansatzes shown in Fig. \ref{ansatze} and perform VQE optimization using two different optimizers: COBYLA and SLSQP from the Qiskit library \cite{Qiskit}. While COBYLA is a gradient-free optimizer, the SLSQP is gradient-dependent, which showed similar performance for a set of selected nuclei across different mass regions \cite{sarma_low-circuit-depth_2026}.  The first panel of Fig. \ref{10B_vqe} shows the convergence to the exact ground state binding energy with the number of iterations corresponding to the COBYLA optimizer.
On the other hand, the second panel shows the same for the SLSQP optimizer. From the figure, it is seen that all three mappings used to represent $^{10}$B ground state that resulted in three different ansatzes reliably converge to the exact diagonalization results from the shell model calculation.

From Fig. \ref{10B_vqe}, it is seen that the SD mapping ansatz converges quickly to the reference energy compared to the other two anstze while using the COBYLA optimizer. On the other hand, for the SLSQP optimizer, while the SD mapping ansatz shows slow convergence compared to the COBYLA counterpart, the convergence for the other two ansatzes is faster as shown in the insets. In order to maintain a consistent parameter choice across all the ansatzes used in this work, we choose optimized parameters from the SLSQP optimizer for ground state energy estimation and sampling the corresponding wavefunctions.

\subsection{Results from noisy simulation and hardware}
Having obtained optimized parameters for each ansatz from the SLSQP optimizer, we transpile those circuits to make them suitable for running on a noisy simulator and on quantum hardware for energy estimation. We choose to run our circuits on IBM FakeFez backend and on $ibm\_fez$ hardware by transpiling our original circuits in terms of the native gates of this quantum hardware (\textit{CZ}, \textit{I}, \textit{Rx}, \textit{Rz}, \textit{Rzz}, $\sqrt{X}$, \textit{X}). Additionally, we performed a level three optimization through the IBM compiler that reduces the gate counts and depth of the transpiled circuits. The level three optimization is the highest optimization stage, applying extensive layout and routing heuristics to circuits already optimized at level two. It also includes the resynthesis of two-qubit blocks to reduce the number of two-qubit operations in the final compiled circuit. The resource counts for the final optimized circuits are shown in Table \ref{rescource}, where "Noisy sim." and "Hardware" represent FakeFez backend and $ibm\_fez$ quantum computer, respectively. Additionally, the number of parameters that are already fixed from the noiseless simulation, the number of Pauli terms, and commuting Pauli groups following the qubit-wise commuting method are also shown.

\begin{figure*}
    \centering
    \includegraphics[width=0.40\linewidth]{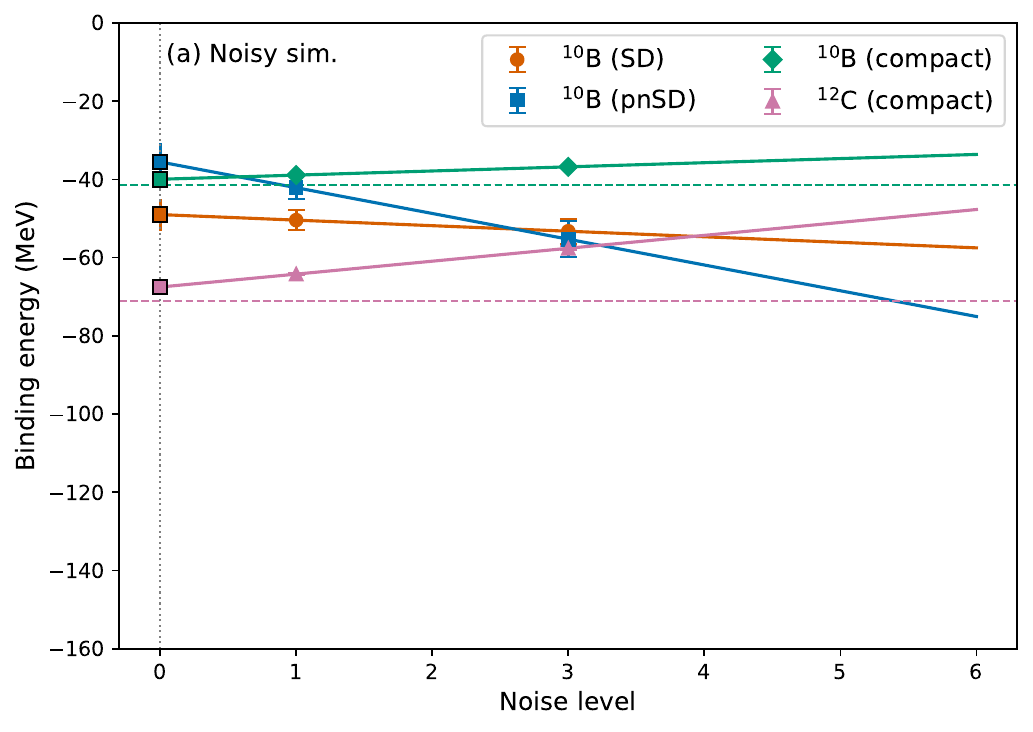}
    \includegraphics[width=0.4\linewidth]{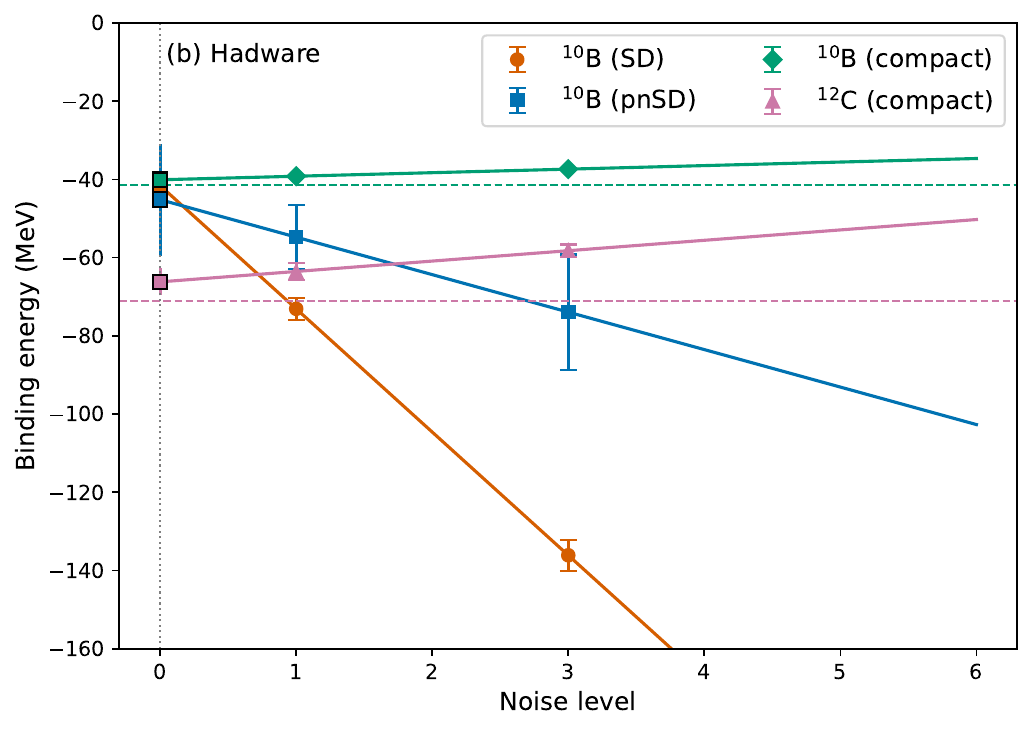}
    \caption{ZNE implemented for (a) noisy simulated results and (b) hardware results.}
    \label{fig_zne}
\end{figure*}

\begin{figure}
    \centering
    \includegraphics[width=\linewidth]{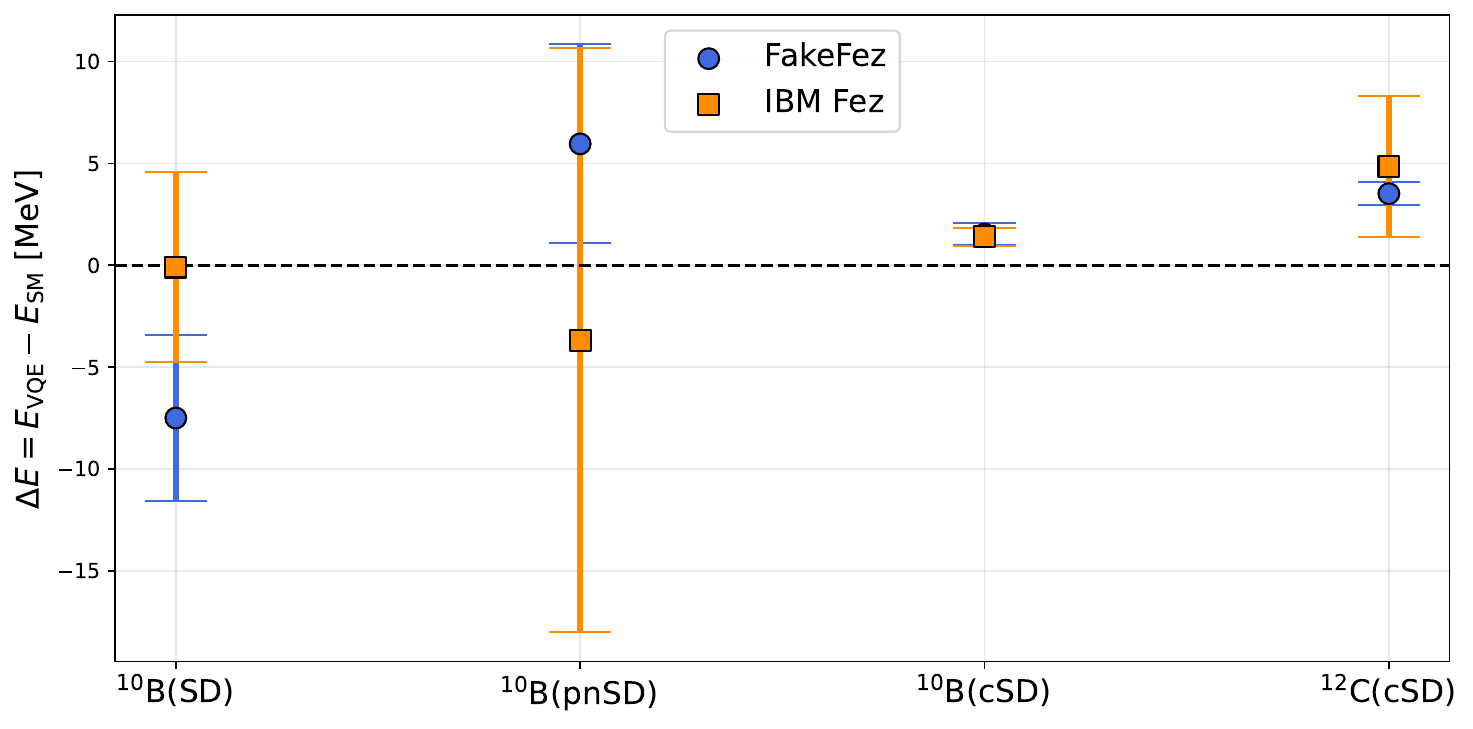}
    \caption{The difference in the ground-state binding energies, $\Delta E$, between the mean error-mitigated results from the noisy simulator and IBM hardware in Table \ref{results} is compared to the shell model results. {\color{black}The error bar represents the uncertainties involved in the extrapolation method.} }
    \label{fig_zne2}
\end{figure}

We first executed the circuits on the FakeFez backend, performing 10 independent runs with 100 shots each. The corresponding results are reported in the “Noise 1” column of Table \ref{results}. We then repeated the procedure on the $ibm\_fez$ hardware, carrying out 5 independent runs with the same number of shots. To mitigate the effects of noise occurring due to the presence of two-qubit gates in the executed circuits in both the noisy simulation and hardware results, we applied zero-noise extrapolation (ZNE). This was done by replacing each $CZ$ gate with a sequence of three $CZ$ gates, noting that a pair of $CZ$ gates acts as the identity. The modified circuits were executed with the same number of independent runs, 10 for the noisy simulation and 5 for the hardware, and the results are presented in the “Noise 3” column of Table \ref{results}. 
Using these two sets of results, we performed a linear extrapolation to estimate the ground-state binding energies at the zero-noise limit, as illustrated in Fig. \ref{fig_zne}. The left panel of Fig. \ref{fig_zne} presents the ZNE procedure applied to the noisy simulation, while the right panel shows the corresponding implementation for the hardware results. 
The final error-mitigated values are listed in the “ZNE” column of Table \ref{results}.
%, with the corresponding percentage errors, relative to the reference energies from shell model calculations reported in the last column. 
{\color{black}As seen in the table, the mean error-mitigated binding energy of the $^{10}$B ground state is -41.593 MeV, obtained from the hardware using SD mapping, and differs from the exact result by only 0.21\%, a substantial improvement over the original deviation of nearly 76\%. However, the uncertainty for this extrapolation is quite large. For the remaining three cases, the mean hardware-based error-mitigated results fall within an 8.88\% error range, whereas the noisy simulation results exhibit larger deviations in most of the cases. In particular, the uncertainty in the extrapolated results from hardware is quite large for pnSD mapping.} 
In Fig. \ref{fig_zne2}, $\Delta E$, which are the differences between extrapolated {\color{black}mean} ground state binding energies obtained from the VQE approach using a noisy simulator and quantum hardware {\color{black}taken from Table \ref{results}}, and the exact results from shell model diagonalization are shown. 
{\color{black}The error bar represents the uncertainties involved in the extrapolation process.
From the figure, it is seen that the mean extrapolated $^{10}$B ground-state binding energy is close to the shell model result for the SD mapping, though the uncertainties are quite large. On the other hand, the mean result for the cSD mapping is also close to the exact results with a little uncertainty.
On the other hand, the pnSD mapping results in a large deviation of around 5 MeV in the mean and also the uncertainty in the extrapolation is quite large. }

\subsection{Wavefunction analysis}
The ansatzes used for SD and pnSD mappings are physics-inspired ones, meaning that each parameter and excitation operator corresponds directly to a specific SD. As a result, in noiseless simulations, these ansatzes naturally reproduce ground state wavefunctions that closely approximate the true wavefunction obtained from the shell model calculation. In contrast, the ansatzes used for cSD mapping are hardware-efficient type, with individual parameters not tied to a specific SD. Consequently, representing the same physical state requires a larger number of parameters than ($N_{SD}$ - 1) for these ansatzes, where $N_{SD}$ is the number of possible SDs for a particular nucleus.

\begin{figure*}
    \centering
    \includegraphics[width=0.85\linewidth]{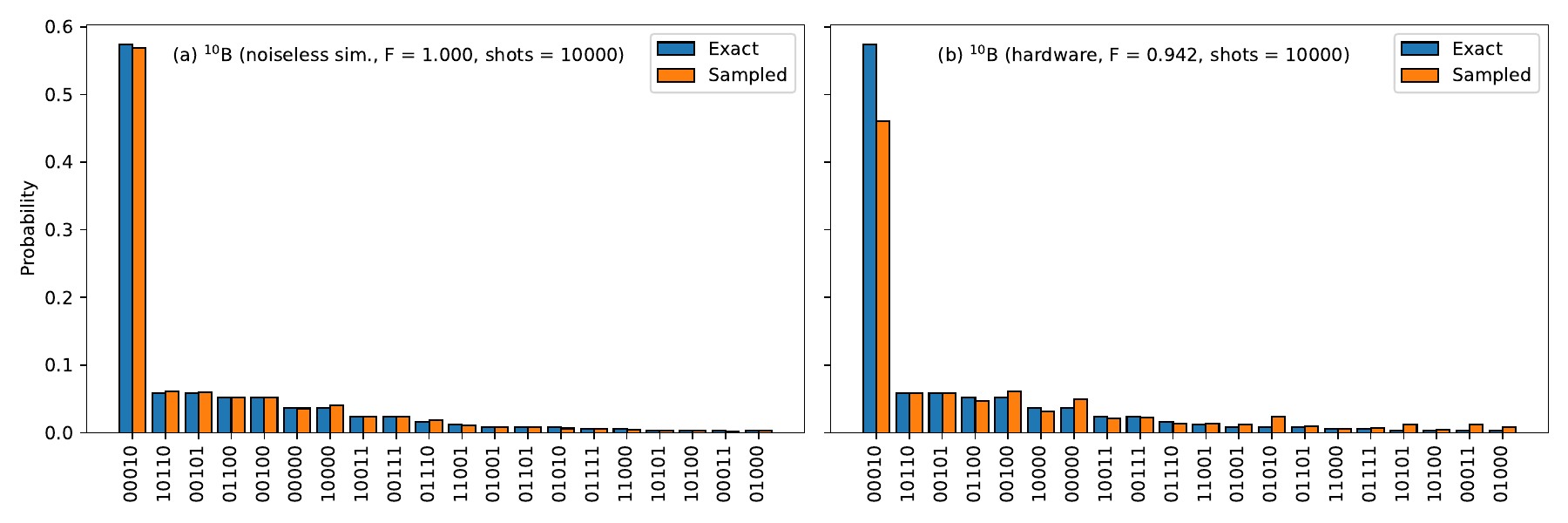}
    \includegraphics[width=0.85\linewidth]{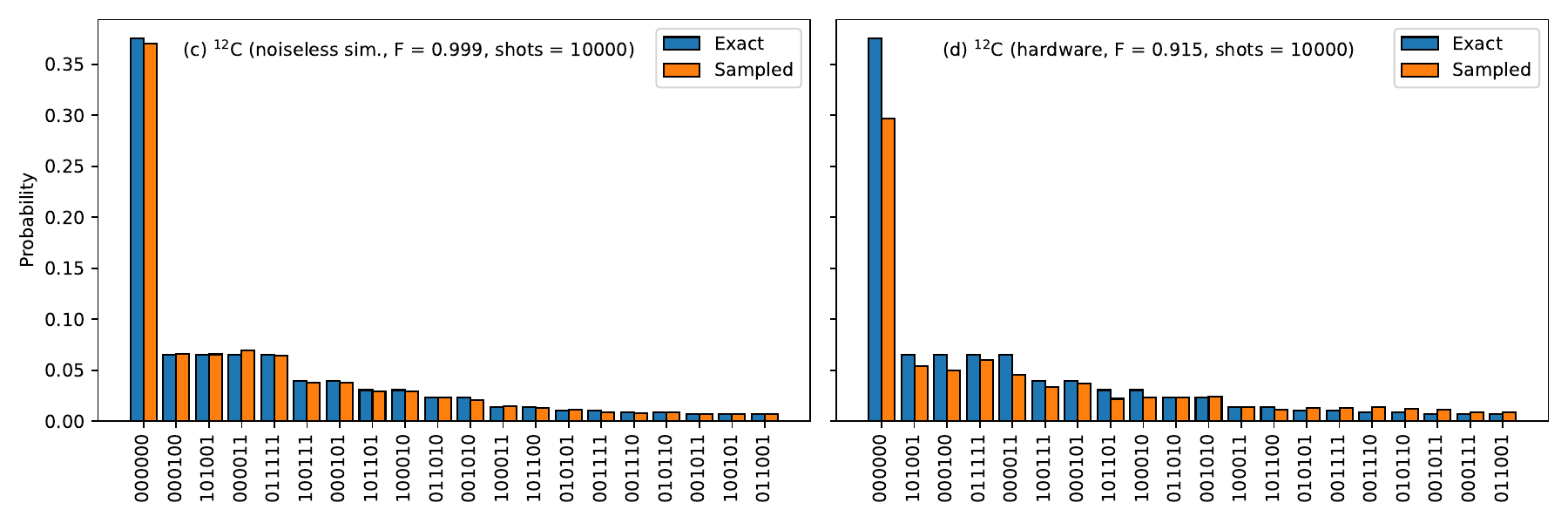}
    \caption{Comparison of the wavefunction obtained from the VQE method following cSD mapping and the exact wavefunction from shell model calculations. Subplots (a) and (b) show the comparison of $^{10}$B ground state from the noiseless simulator and $ibm\_fez$ hardware. Subplots (c) and (d) show the same comparison for the ground state of $^{12}$C.}
    \label{fig_wf}
\end{figure*}

The 5-qubit cSD ansatz for $^{10}$B ground state requires six layers of $CNOT$ plus rotation gates having 35 parameters to converge to exact results. On the other hand, the 6-qubit cSD ansatz for 
{\color{black}the $^{12}$C} 
ground state requires 10 layers of $CNOT$ plus rotation gates having 66 parameters for its convergence to the exact binding energy. After getting the optimized parameters for these ansatzes from the SLSQP optimizer, we transpile those circuits to make them suitable for sampling the wavefunctions on a noiseless simulator. The subplots (a) and (c) of Fig. \ref{fig_wf} show a comparison of the exact wavefunctions from the shell model calculation and the VQE wavefunctions from noiseless simulation following cSD mapping for $^{10}$B and $^{12}$C ground state. We calculate the wavefunction fidelity, $F$ = $|\langle \psi_{SM}|\psi_{VQE}\rangle|^2$, where $\psi_{SM}$ and $\psi_{VQE}$ represent the shell model wavefunction and the VQE wavefunction obtained from sampling on a noiseless simulator. For a sampling with 10000 shots, the noiseless simulator provides wavefunctions close to $F$ = 1, meaning that the $\psi_{SM}$ and $\psi_{VQE}$ are in excellent agreement. We repeat the same procedure for $ibm\_fez$ hardware and compare to the shell model results in subplots (b) and (d) of Fig. \ref{fig_wf} for the same two nuclei. The fidelities of the wavefunction sampled on hardware with 10000 shots are  $F$ = 0.942 for $^{10}$B ground state and $F$ = 0.915 for $^{12}$C ground state. For both the cases, a major difference is observed for the most dominant SDs: '00010' for $^{10}$B that corresponds to $|1, 2, 3; 7, 8, 9\rangle$ and '000000' for $^{12}$C that corresponds to $|0, 1, 2, 3; 6, 7, 8, 9\rangle$. It should be noted that the wavefunctions obtained from the hardware have not undergone any error-mitigation techniques. We can expect improved fidelities with the implementation of error-mitigation techniques like the evaluation of the binding energies. {\color{black}The codes and data presented in this work are available in the GitHub repository \cite{code_repo}.}

\section{Discussion}
Out of the three mappings considered in this work for determining the ground state of $^{10}$B, the compact SD (cSD) mapping uses qubits most efficiently, but this comes at the cost of a larger number of variational parameters. It also requires fewer gates to construct the trial wavefunction than the other two mappings. In contrast, the SD mapping is more efficient for energy estimation on hardware in terms of execution time because its Hamiltonian can be grouped into only four commuting Pauli groups. As a result, the circuits need to be executed only four times to estimate the energy, once for each Pauli group. This advantage would have been even more significant if the intermediate VQE energy evaluations had also been performed on hardware rather than on a noiseless simulator. However, the trial wavefunctions in the SD mapping require a moderately larger number of gates and circuit depth compared to the cSD mapping. The pnSD mapping, meanwhile, requires deeper circuits than both cSD and SD, although it uses fewer qubits than the SD mapping. Estimating the ground-state energy with pnSD on noisy simulators or real hardware is particularly expensive because the deep circuit, which contains more than 300 two-qubit gates, must be executed 530 times due to the Hamiltonian being divided into 530 commuting Pauli groups.

Representing the ground state (ground state) of $^{12}$C using the SD and pnSD mappings would require 51 and 30 qubits, respectively, making VQE optimization on a noiseless simulator computationally challenging. As a result, we present results only for the cSD mapping for this nucleus. The cSD mapping, when combined with a hardware-efficient ansatz, has the potential to describe a broader range of nuclei across different mass regions, particularly in cases where the construction of a UCC 
{\color{black}or ADAPT-VQE type ansatz is challenging in terms of implementation on noisy quantum hardware. For example, the implementation of ADAPT-VQE for ${}^{10}$Be requires 12 qubits and approximately 48 layers, having an average of 62 2$Q$ gates per layer, requiring 2976 $2Q$ gates \cite{perez-obiol_nuclear_2023}. Since the number of Slater determinants (SDs) is 51 for both ${}^{10}$Be and ${}^{12}$C, we can roughly assume that approximately 3000 2$Q$ gates would be required to implement the ADAPT-VQE ansatz for ${}^{12}$C. In contrast, the cSD mapping reported in this work requires a variational ansatz involving 6 qubits and only 50 2$Q$ gates. The reduction in the required quantum resources for cSD mapping is quite similar to the Gray code encoding implemented for nuclear systems \cite{nifeeya, PhysRevA.103.042405, PhysRevC.104.034301}.}
{\color{black}The Gray-code mapping encodes the many-body basis states using a Gray-code representation and constructs the corresponding Pauli-string Hamiltonian through a generalized operator transformation \cite{nifeeya}, thereby exploiting the structure of transitions between basis states. In contrast, the cSD mapping directly embeds the shell-model Hamiltonian matrix into the nearest 2$^n$-dimensional qubit Hilbert space and decomposes it into Pauli strings, achieving comparable qubit compression with a simpler, basis-independent construction.}
% {\color{black}For cSD mapping, the Pauli string Hamiltonian is obtained from a Hamiltonian matrix defined in a many-particle basis where fermionic antisymmetrization is already taken into account. The Gray code encoding, on the other hand, converts the individual terms of the second-quantized Hamiltonian into Pauli strings.} 
Finally, from the wavefunction analysis, it is seen that although the cSD mapping requires a larger number of parameters to define the ground state exactly, it can still generate a high-fidelity wavefunction from quantum hardware. 
{\color{black}It is to be noted that in Ref. \cite{17n2-xh6k}, a similar set of $p$-shell nuclei were studied within the VQE approaches based on the total number of operations needed for complete optimization. This quantity could be highly dependent on the choice of optimizer and how many energy evaluations are required per iteration. As we are using two different optimizers, SLSQP and COBYLA, which are different from the L-BFGS-B optimizer used in Ref. \cite{17n2-xh6k}, a direct comparison is not shown in this work.}

\section{Summary and conclusion}
In this work, we use the VQE algorithm to study the ground state of two mid $p$-shell nuclei: $^{10}$B and $^{12}$C. Based on how many-body configurations are mapped into qubits, we tested three different mappings for the ground state of  $^{10}$B. The SD mapping requires the highest number of qubits, 26, followed by the pnSD mapping requiring 20 qubits to define the ground state On the other hand, the cSD mapping requires only 5 qubits to define the same problem. For each mapping, we construct the Hamiltonians and trial wavefunctions separately, and they are compared in terms of resource counts. Although the SD mapping requires a large number of qubits, its execution on quantum hardware requires less execution time, as the full Pauli string Hamiltonian can be decomposed into only four commuting Pauli groups. On the other hand, the qubit-efficient cSD mapping requires fewer gates to construct the trial wavefunction at the expense of a larger number of parameters. Finally, the pnSD mapping requires deep circuits with a slightly reduced number of qubits compared to the SD mapping. However, the pnSD circuit is expensive to run on noisy simulators and hardware, as the Hamiltonian comprises a larger number of commuting Pauli groups. 
{\color{black}The mean binding energy obtained from the $ibm\_fez$ hardware following ZNE) for the SD mapping differs from the exact shell-model result by 0.21\%, although the associated uncertainty is large. In comparison, the cSD mapping yields a 3.37\% deviation with a smaller uncertainty. These percentage deviations represent the residual differences from the reference energy and should not be interpreted as measures of the total simulation uncertainty.} 
We then showed the potential of this cSD mapping to define the ground state of another mid-$p$-shell nucleus, $^{12}$C, where SD and pnSD mapping become computationally expensive for a normal classical computer 
{\color{black}following a statevector simulation where other mappings could be more effective.} 
Finally, we showed the fidelity of the VQE wavefunctions following cSD mapping compared to the exact shell model wavefunctions for both the nuclei considered in this work. 

In conclusion, all three mappings considered in this work have certain advantages over the others. The cSD mapping holds the promise to deal with complex nuclear systems in different mass regions in terms of qubit efficiency for the NISQ era quantum hardware. Similarly, this cSD mapping can be used to target excited states as well, following constrained VQE versions with a similar structure of the trial wavefunction. On the other hand, SD and pnSD mapping cases could be interesting for future quantum devices, and also for the initial state preparation for other quantum algorithms, where each possible SD can be controlled by a single parameter.

\section*{Acknowledgements}
This work is funded and supported by UK STFC grant ST/Y000358/1 and by the UK National Quantum Computer Centre [NQCC200921], which is a UKRI Centre and part of the UK National Quantum Technologies Programme (NQTP).
\appendix

\section{Excitation gates in terms of basic gates}

\begin{figure*}
    \centering

    \textbf{(a)}\\
    \includegraphics[width=0.30\linewidth]{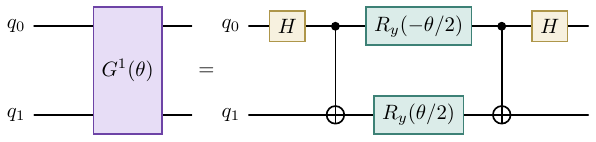}

    \vspace{2.5mm}

    \textbf{(b)}\\
    \includegraphics[width=0.50\linewidth]{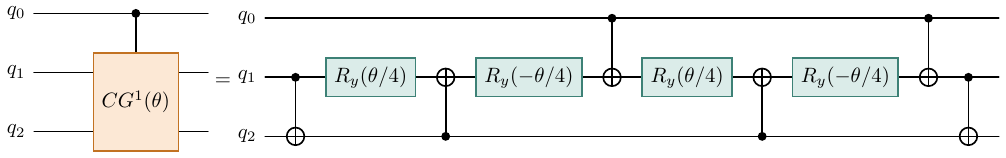}

    \vspace{2.5mm}

    \textbf{(c)}\\
    \includegraphics[width=0.85\linewidth]{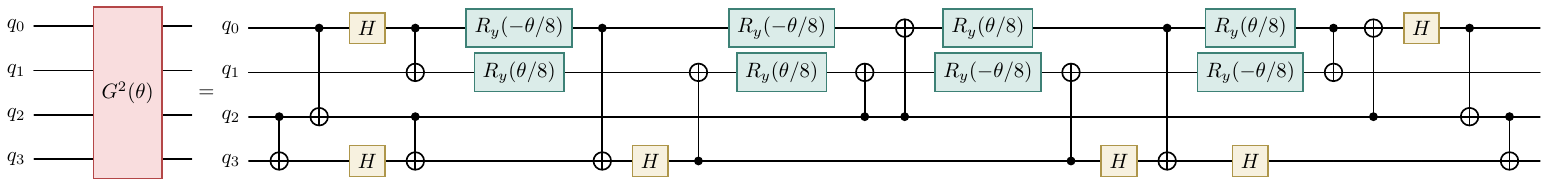}

    \caption{(a) Single excitation gate ($G^1$) in terms of ($H$, $R_Y$, and $CNOT$) gates adapted from \cite{single_ex}, (b) controlled single excitation gate ($CG^1$) in terms of ($R_Y$, and $CNOT$) gates adapted from \cite{sarma_prediction_2023}, (c) double excitation gate ($G^2$) in terms of ($H$, $R_Y$, and $CNOT$) gates adapted from \cite{double_ex}.}
    \label{Double_ex}
\end{figure*}

Following the SD mapping, the trial wavefunction for the ground state of $^{10}$B is constructed by repeated application of single excitation $G^1$ operators. This excitation gate can be decomposed into ($H$, $R_Y$, and $CNOT$) gates as shown in the first panel of Fig. \ref{Double_ex}. On the other hand, following the pnSD mapping involves applying controlled single excitation gates ($CG^1$), which can be decomposed into ($R_Y$ and $CNOT$) as shown in the second panel of Fig. \ref{Double_ex} and double excitation gates ($G^2$) that can be further decomposed into a set of ($H$, $R_Y$, and $CNOT$) as shown in the third panel of Fig. \ref{Double_ex}.

\bibliographystyle{unsrtnat}
\bibliography{ref}



%% For example:

%% \bibitem[Aladro et al.(2015)]{Aladro15} Aladro, R., Martín, S., Riquelme, D., et al. 2015, \aas, 579, A101

%%\end{thebibliography}

\end{document}